\documentclass[journal]{IEEEtran}

\IEEEoverridecommandlockouts

\usepackage{cite,fancyhdr}
\usepackage{amsmath,amssymb,amsfonts}
\usepackage{algorithmic}
\usepackage{graphicx}
\usepackage{textcomp}
\usepackage{xcolor}
\def\BibTeX{{\rm B\kern-.05em{\sc i\kern-.025em b}\kern-.08em
    T\kern-.1667em\lower.7ex\hbox{E}\kern-.125emX}}

\usepackage{balance}

\newcommand{\equals}{\!=\!}

\newcommand{\gteq}{\!\ge\!}

\newcommand{\lthan}{\!<\!}

\DeclareMathOperator{\argmax}{argmax}

\begin{document}

\title{Federated Learning for Remote IoT over LPWAN: Preliminary Insights From Structural \\Health Monitoring

\thanks{
This work was supported by the Anusandhan National Research Foundation (ANRF), Government of India, through its Advanced Research Grant (ARG) No. ANRF/ARG/2025/011639/ENS.}
}

\author{Lakshmiprasanna Bandaru, Anshika Singh, Tarutal Ghosh Mondal, \\T. Jothi Saravanan, and Siddhartha S. Borkotoky \\ \textit{Indian Institute of Technology Bhubaneswar, Khordha 752051, Odisha, India}\\
\texttt{24sp06005@iitbbs.ac.in}
}

\maketitle
\thispagestyle{fancy}
\lhead{\textit{Accepted for Presentation at IEEE ANTS 2026.}}

\begin{abstract}
Realizing federated learning (FL) in remote regions lacking reliable cellular infrastructure is challenging, as it requires dependable communication among participating devices. Low-power wide-area networks (LPWANs) offer an attractive communication solution for such scenarios. However, their limited data rates  necessitate careful investigation of the communication-learning trade-offs and mechanisms to reduce communication overhead. This paper presents an investigation of the Long Range (LoRa) LPWAN technology for FL-based structural health monitoring (SHM), a representative remote IoT application.  We consider a SCAFFOLD-based FL framework for multi-class anomaly detection and analyze the latency-accuracy trade-off. We also examine strategies for allocating the LoRa spreading factor (SF) during model-update exchange. Numerical results show that, with appropriate SF allocation, LoRa-supported FL-SHM can achieve competitive learning accuracy. In particular, jointly accounting for link distances and clients' dataset sizes when selecting the SF provides a favorable balance between convergence performance and communication latency. These preliminary findings offer useful insights into the design of FL systems over LPWANs for remote IoT deployments.
\end{abstract}


\section{Introduction}
Federated learning (FL) has emerged as a promising paradigm for enabling collaborative machine learning across geographically distributed Internet of Things (IoT) devices. FL allows clients to collaboratively train a machine-learning model through iterative exchange of model parameters rather than raw data, thus preserving data privacy. However, realizing FL in remote regions lacking reliable cellular infrastructure remains challenging, as it requires reliable communication among participating devices. Legacy wireless technologies such as Bluetooth, WiFi, and ZigBee cannot support the long communication ranges typically encountered in such deployments without multi-hop solutions, which increase system complexity. Cellular technologies provide an alternative; however, their reliance on communication infrastructure limits deployment in regions lacking coverage. Low Power Wide Area (LPWA) technologies offer a promising solution by enabling long-range communication at low power. However, these benefits come at the cost of low data rates and interference-induced packet losses due to operation in unlicensed spectrum, whereas FL requires dependable iterative transfer of substantial  amounts of information among devices.

An investigation of this trade-off in the context of remote IoT deployments is the focus of this work. For this purpose, we consider the representative use case of population-based structural health monitoring (SHM), where sensor data from multiple geographically distributed structures are utilized for learning-aided analysis, fault detection, and predictive maintenance~\cite{BGG21}. In view of the privacy and security concerns associated with the sharing of structure-specific data for centralized learning, FL provides a natural solution for learning-driven SHM. This has led to a recent surge of interest in FL-aided SHM~\cite{CZC25,AA23,YLQ25,ASN21,PTH25,PZZ24}. However, structures such as bridges are often located in remote regions with limited cellular infrastructure. This motivates us to investigate the latency-accuracy trade-off of an FL-based SHM system supported by bandwidth-constrained LPWANs. 

\begin{figure}[!t]
    \centering
    \includegraphics[width=1\linewidth]{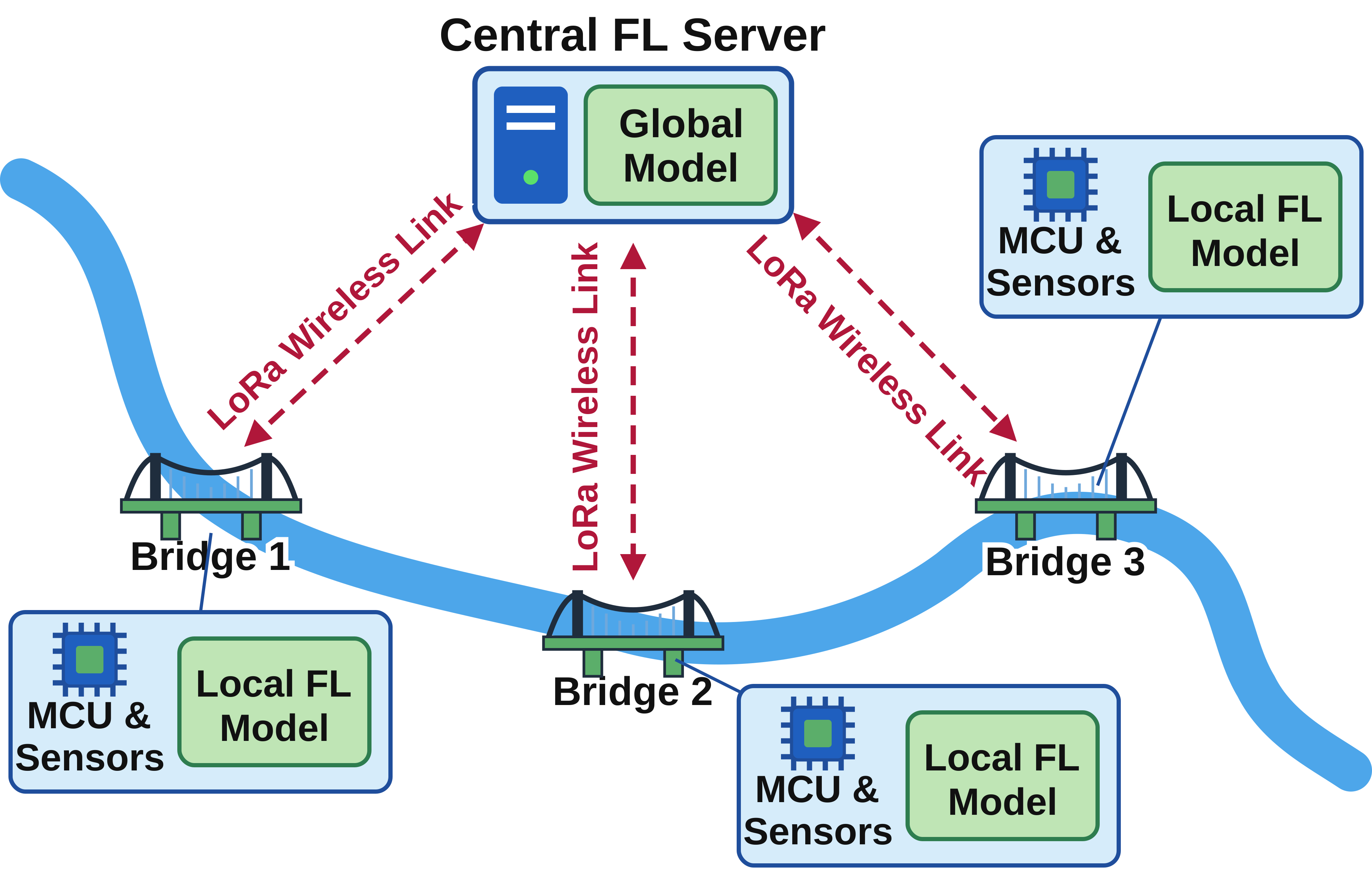}
    \caption{Representative overview of LoRa-based FL for remote SHM.}
    \label{fig:SHM_diagram}
\end{figure}

Specifically, we consider Long Range (LoRa) communications, a prominent LPWA technology, for FL parameter exchange in the context of a population-based SHM framework that employs SCAFFOLD-based anomaly detection using a cable-stayed bridge dataset. A representative depiction of such a scenario is shown in Fig.~\ref{fig:SHM_diagram}. For the considered setup, we characterize the learning accuracy, training time, and device airtime. We further investigate spreading-factor (SF) allocation strategies for LoRa-enabled FL and demonstrate the benefits of jointly accounting for communication conditions and client data contribution during SF selection. The resulting observations provide useful preliminary insights into the design of FL systems over LPWANs for remote IoT deployments.

\subsection{Related Works}

FL-based SHM has recently received significant attention. A clustered FL framework for population-based structural health monitoring across multiple structures is proposed in \cite{CZC25}, while \cite{AA23} presents a personalized FL approach that incorporates tensor-based data fusion for improved damage detection. In~\cite{YLQ25}, the authors propose an FL-based SHM mechanism for monitoring miter gates that are used for navigation and in flood control systems. An FL model augmented with tensor data fusion to detect structural damages is proposed in~\cite{ASN21}. A federated convolutional neural network model with a data standardization procedure for damage identification in submerged structure-foundation systems is developed in~\cite{PTH25}. In~\cite{PZZ24}, FL is used to detect  damages in gantry crane structures. These works focus on the development of FL models for SHM under the assumption of an ideal communication channel. In practice, however, communication reliability and bandwidth limitations have a profound impact on the performance of the FL algorithm. Packet losses on the communication link impact the FL model's accuracy, and communication bandwidth impacts training time by dictating the airtime needed to transfer updates.  To investigate the impact of bandwidth-constrained and erasure-prone wireless links on FL-SHM, we examine the integration of an LPWA layer to an FL-based SHM application.


\subsection{Contributions of This Paper}
Our main contributions are as follows:

\begin{enumerate}
    \item We investigate a Stochastic Controlled Averaging for Federated Learning (SCAFFOLD) system for multi-class sensor anomaly detection using a real cable-stayed bridge dataset.  
    We evaluate the accuracy of the system when updates are transferred over LoRa links.   

    \item We quantify the accuracy degradation due to lossy quantization and sparsification, which are necessary to reduce update sizes for LoRa's low data rate constraints. Specifically, we examine the impact of a three-stage compression pipeline combining top-$k$ sparsification, int8 quantization, and zlib lossless encoding. 

    \item We further investigate and quantify the additional accuracy penalty due to LoRa frame loss enforced by signal attenuation and interference.  
    
    \item We investigate spreading factor (SF) allocation for update transmission. Three strategies are devised: (1) based on the farthest client, (2) based on the client with the largest dataset, and (3) based on a weighted average global model update delivery probability, with weights determined by dataset sizes. We show that for the evaluated scenario, Strategy 3 provides the best performance among the three in terms of both accuracy and model transmission time.  
\end{enumerate}

\section{Federated-Learning Framework}

\subsection{Dataset and Model Configuration }

We employ a dataset collected from a long-span cable-stayed bridge located in China~\cite{BLN21}. The SHM objective is treated as a multi-class classification problem based on images of vibration signal patterns. The dataset contains seven classes of sensor anomaly behaviors, namely: normal, missing, minor, outlier, square, trend, and drift. Each class is available at a unique client, resulting in seven clients and a strictly non-independent and identically distributed (non-IID) data split among clients. Furthermore, there is large variation among the clients in terms of dataset size, which ranges from 527 samples to 13,575 samples.  A deep learning model based on the InceptionV3 architecture~\cite{SVI16} is used for classification. Before training, the input data is resized and normalized to ensure stable convergence.

\subsection{Federated Learning Framework }
We implement a centralized FL framework with one central server and seven clients. Training occurs over several communication rounds, as explained below:

\begin{enumerate}
    \item \textit{Initialization:} The central server initializes a global model with random parameters.
    
    \item \textit{Model Distribution:} In each communication round, the server sends the current global model to all participating clients.
    
    \item \textit{Local Training:} Each client trains the received model on its local dataset and updates the model parameters.
    
    \item \textit{Update Transmission:} The updated parameters from each client are sent back to the server.
    
    \item \textit{Aggregation:} The server aggregates the received updates to form an improved global model.
    
    \item \textit{Iteration:} The updated global model is shared again with the clients, and this process continues until the model converges or reaches the desired performance.
\end{enumerate}

\subsection{SCAFFOLD-Based Aggregation}

The non-IID distribution of data across clients can result in \textit{client drift}. Because each client has data from only one class, their local updates tend to move in different directions. This impedes convergence and reduces overall model performance. Traditional methods like Federated Averaging (FedAvg) are particularly affected, as they simply average these different updates. To address this issue, we use the Stochastic Controlled Averaging for Federated Learning (SCAFFOLD) algorithm~\cite{KKM20}. SCAFFOLD introduces \textit{control variates} at both the server and client sides, which act as correction terms to guide local training toward the global objective.

In each communication round, the server sends the global model along with its control variate to all clients. Each client then trains the model on its local data while applying a correction term based on the difference between the server and client control variates. After training, clients send back their updated model parameters and control variates. The server aggregates these updates and adjusts its control variate to produce an improved global model. By reducing client drift, SCAFFOLD enables more stable training and faster convergence than FedAvg, especially in this highly non-IID setting where each client holds data from only one class.

Let $c$ and $c_i$ denote the server and client-$i$ control variates, respectively, and let $g_i$ denote the local mini-batch gradient. During local training, client $i$ updates its model using the corrected gradient $g_i+c-c_i$. After local training, the client returns its model change
$\Delta w_i$ and control-variate change $\Delta c_i$. Let $\mathcal{S}$ denote
the set of clients whose updates are successfully received by the server. The
server updates the global model and its control variate as follows~\cite{KKM20}:
\begin{align}
w &\leftarrow w+\eta_g\frac{1}{|\mathcal{S}|}\sum_{i\in\mathcal{S}}\Delta w_i, \\
c &\leftarrow c+\frac{1}{N}\sum_{i\in\mathcal{S}}\Delta c_i,
\end{align}
where $\eta_g$ is the server step size and $N$ is the total number of clients.

\section{Communication Framework}
\label{sec:commm_framework}
The clients and the server use LoRa communications for exchanging model updates. To ensure the feasibility of transmission over bandwidth-constrained LoRa links, the update is compressed prior to transmission via the following stages:

\begin{enumerate}
    \item Top-$k$ sparsification: Only those elements with absolute values above a threshold are transmitted. The rest are discarded, which reduces the number of parameters sent in each round. In our implementation, only the top 0.1\% of the total parameters are transmitted per round.

    \item Int8 quantization: The selected model parameters are converted from 32-bit floating-point (float32) values to 8-bit integers (int8). This reduces the storage and transmission cost per parameter by a factor of 4.

    \item Zlib Compression: The sparsified and quantized updates are further compressed using the zlib lossless compression algorithm. This step leverages redundancy in the data to reduce its size without losing any information. 
\end{enumerate}

The compressed update is transmitted using LoRa. Since the update is too large to fit in the payload of a single LoRa frame, it is divided into smaller fragments of size not exceeding LoRa's maximum transmission unit. On the uplink (i.e., from a client to the server), the transfer of the client's local update is modeled as point-to-point unicast communication. On the downlink, the server broadcasts the updated global model to all seven clients.

The key physical layer parameter that dictates LoRa's communication performance is the SF~\cite{GeR17}, which takes integer values between 7 and 12. Increasing the SF improves receiver sensitivity, resulting in a longer communication range. However, this also incurs  longer transmission times and energy
consumption. Mechanisms for selecting the SF are discussed in Section~\ref{sec:SF_allocation}.

\subsection{Forward Erasure Correction (FEC) and Model Delivery Probability (MDP)}
To compensate for frame losses that occur on the wireless links, an application-layer forward erasure correction (FEC) mechanism is used. Let $k$ be the number of source fragments comprising the compressed update to be sent. A rate-$r$ FEC ($r\lthan 1$) produces $n \equals k/r$ coded fragments, each of which is a linear combination of the source fragments. Each coded fragment is transmitted as a single LoRa frame. The receiver is able to reconstruct the update as long as $k$ linearly independent coded fragments are received. We denote the probability of this event as the \textit{model delivery probability} (MDP). If a maximum-distance separable code (MDS) is used for producing coded fragments, then the destination can reconstruct the update upon receiving any $k$ coded fragments. For other types of encoding, slightly more than $k$ successful receptions may be needed. For analytical tractability, we assume an MDS code in this work. Thus, the MDP for a source-destination link is the probability that at least $k$ out of the $n$ transmitted LoRa frames are successfully received. Letting $d_i$ denote the distance between the server and client $i$, the MDP when SF $j$ is used for transferring the update between the server to client $i$ or vice versa is 
\begin{align}
    \mathcal{P}_j(i) = \sum_{l = k}^{n} {n \choose l}\left(\mathcal{S}_j(d_i)\right)^l  \left(1 - \mathcal{S}_j(d_i)\right)^{n-l},
\end{align}
where $\mathcal{S}_j(d_i)$ denotes the probability that a LoRa frame transmitted over a link of length $d_i$ is successfully delivered. Methods to find $\mathcal{S}_j(d_i)$ can be found, for example, in~\cite{GeR17} and~\cite{MSG19}.

\subsection{Spreading Factor Allocation}
\label{sec:SF_allocation}
As mentioned earlier, there is an inherent tradeoff between LoRa's communication reliability and transmission airtime. For FL-SHM, this translates to a tradeoff between the model accuracy and the time needed to train the model. Consequently, achieving an acceptable model accuracy within reasonable time requires careful assignment of the SF. To assess the impact of the SF selection, we investigate three strategies. The strategies employ a design parameter $\delta$ defined as the \textit{target delivery probability}. The strategies are described below:

\subsubsection{Farthest-Client Based} This strategy selects the smallest SF that provides an MDP of  $\delta$ at all clients. This is equivalent to ensuring such an MDP at the client with the worst MDP, which typically is the client farthest from the server. In general, SF $j^*$ is chosen if
\begin{align}
    j^* = \min \{j | \mathcal{P}_j(i') \gteq \delta\},
\end{align}
where 
$i' \equals \argmax_i\{d_i\}$.

\subsubsection{Largest-Dataset Based} The farthest-client strategy aims to provide coverage for all clients but may result in a higher SF than necessary if the farthest client contributes little to the global model. As an alternative, the largest-dataset strategy chooses the SF  based on the client holding the largest number of training samples, since this client contributes most to federated aggregation. Specifically, the smallest SF that provides a MDP of $\delta$ or above at this client is chosen for transmissions. That is, SF $j^*$ is chosen if
\begin{align}
    j^* = \min \{j | \mathcal{P}_j(i'') \gteq \delta\},
\end{align}
where 
\begin{align}
i'' = \argmax_i\{T_i\},
\end{align}
where $T_i$ denotes the number of training samples at client $i$.

\subsubsection{Weighted-Average-MDP Based}
While the farthest-client strategy may choose an unnecessarily large SF, the largest-dataset approach may inhibit the participation of clients with smaller datasets in the FL process. This may be detrimental in a non-IID setting like ours. To strike a balance between the two approaches, we consider a third strategy that jointly considers communication reliability and data contribution by computing a weighted average of the MDPs across all clients, with the weighting factor proportional to the sample size. Specifically, we define the weighted average MDP for SF $j$ as
\begin{align}
    \overline{\mathcal{P}}_j = \frac{1}{T_{\mathrm{tot}}}\sum_{i=1}^{N} \mathcal{P}_j(i) T_i,
\end{align}
where $T_{\mathrm{tot}}$ is the total number of training samples across all clients, given by 
\begin{align}
    T_{\mathrm{tot}} = \sum_{i=1}^{N} T_i.
\end{align}

SF $j^*$ is chosen for communication if
\begin{align}
    j^* = \min \{j | \overline{\mathcal{P}}_j \gteq \delta\}.
\end{align}

Under the imbalanced dataset distribution considered here, this method assigns greater importance to clients with larger datasets while still considering the link quality of all clients.

\section{Simulation Setup}
We consider random uniform placement of the clients within a circular area of radius 500 m. The server is located at the center. An FL session consists of 30 communication rounds. The presented results are the average of 10 such sessions, each with random assignment of client locations. At the start of each session, the InceptionV3 backbone is loaded with its pretrained ImageNet weights and kept frozen. Only the trainable classification head, consisting of a 1024-unit dense layer followed by a 7-class softmax output layer, is randomly initialized. The resulting global model is then copied to every client, ensuring that all clients start from the same initial state. The parameters of the classification head are trained locally and exchanged between the clients and the server during federated learning. Each client uses stochastic gradient descent (SGD) with a learning rate of 0.01, no momentum, a batch size of 64, and one local epoch per communication round. The server step size is $\eta_g \equals 0.01$, and the server and client control variates are initialized to zero. All seven clients participate in every round; a client that fails to receive the global model in a round does not perform local training in that round. The dataset is divided into 90\% for training and 10\% for testing. Model accuracy is evaluated on the test set. 

For transmitting model updates, the transmission power is fixed at 14 dBm for all devices. The links are subject to Rayleigh fading and have path loss exponent of 3. A LoRa frame may be lost either due to interference or because its received power is below the receiver sensitivity. The receiver sensitivity values for each SF can be found in~\cite{GeR17}. The strongest interferer model~\cite{GeR17} with capture thresholds given in~\cite{MSG19} is used for modeling interference-induced losses. The interferers are assumed to be distributed according to a Poisson Point Process of intensity $1\!\times\! 10^{-4}$ devices per square meter.     
For erasure correction, an FEC of rate $r \equals 0.5$ is applied.

\section{Numerical Results}
\begin{figure}[!t]
\centering
\includegraphics[width=\columnwidth]{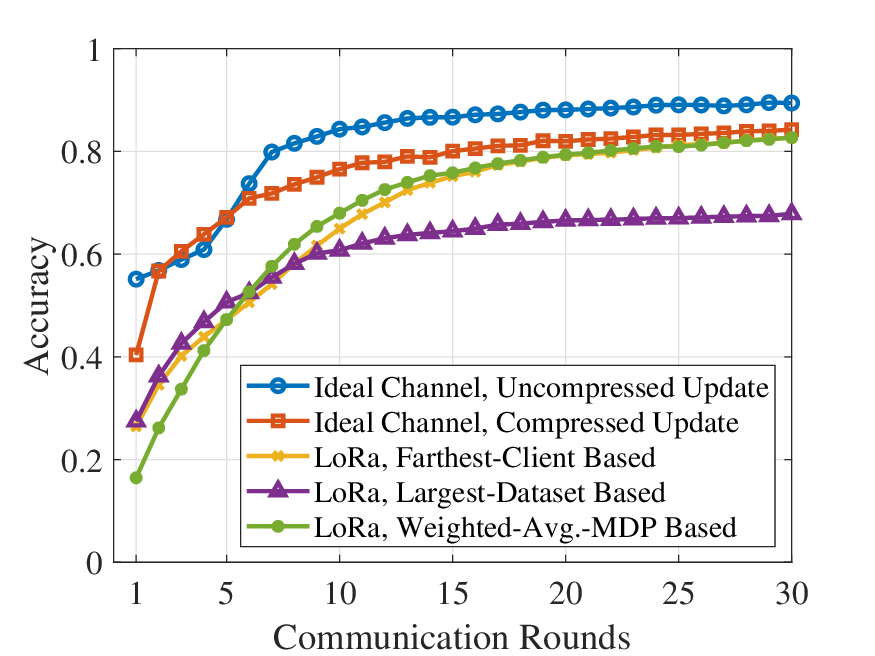}
\caption{Accuracy vs. communication rounds.}
\label{fig:1}
\end{figure}

Fig. \ref{fig:1} shows the evolution of model accuracy across communication rounds under the three SF-allocation mechanisms. The results are compared against two benchmark schemes, described below.  

\textit{Benchmark~1 (Ideal Channel, Uncompressed Update):} This is a hypothetical scheme in which uncompressed updates are exchanged over error-free channels. Clearly, this is the most favorable situation for FL, as it suffers no update loss from lossy compression or from frame erasures.  

\textit{Benchmark~2 (Ideal Channel, Compressed Update):} Here, we retain the ideal channel but relax the assumption of lossless compression. Instead, we apply the compression pipeline of Section~\ref{sec:commm_framework} comprising top-$k$ sparsification, int8 quantization, and zlib compression. The size of the raw update in float32 format is approximately 8,420~kB. The size of the updates after each compression stage is as follows:  After top-$k$ sparsification, which retains 2106 entries
(top 0.1\%), the update reduces to 16.8~kB. The reduced size accounts for the retained entries and their corresponding identifiers. After int8 quantization, the size reduces to 10.6~kB, and after zlib compression to approximately 5~kB (4.8--5.1~kB across clients). Note that these are the same update sizes handled by the practical LoRa-based schemes.

As seen from Fig~\ref{fig:1}, \mbox{benchmark~1} outperforms benchmark~2 owing to the transmission of uncompressed updates, converging to about 90\% accuracy as opposed to benchmark~2's 85\% accuracy. The slight under-performance of benchmark~1 in the early rounds is attributed to random model initialization. 

For the practical LoRa-based methods, we observe that their performance lags that of the hypothetical benchmarks. This is due to the frame losses in a  practical communication system.  However, by round 20, the practical FL systems employing farthest-client and weighted-average-MDP based SF selection come very close to the performance of the hypothetical benchmark~2 that employs compressed update transmission (reaching about 83\% accuracy as opposed to benchmark~2's 85\%). By contrast, largest-dataset-based SF allocation provides relatively poor performance, saturating at about 68\% accuracy. In summary, the results demonstrate that, with appropriate SF selection, it is possible for a LoRaWAN-aided SHM to reach accuracies close to a hypothetical system with error-free communications.  

Among the practical SF-allocation strategies, largest-dataset-based selection provides lower FL accuracy because it selects the SF based only on the client with the most data, ignoring whether other clients can actually receive the model. Since every client in our setup represents a unique anomaly class, missing even one client during aggregation hurts the overall model quality. The farthest-client-based and average-weighted-MDP-based strategies avoid this shortcoming, since the former covers all clients whereas the latter jointly considers data size and link quality.

\begin{figure}[!t]
\centering
\includegraphics[width=\columnwidth]{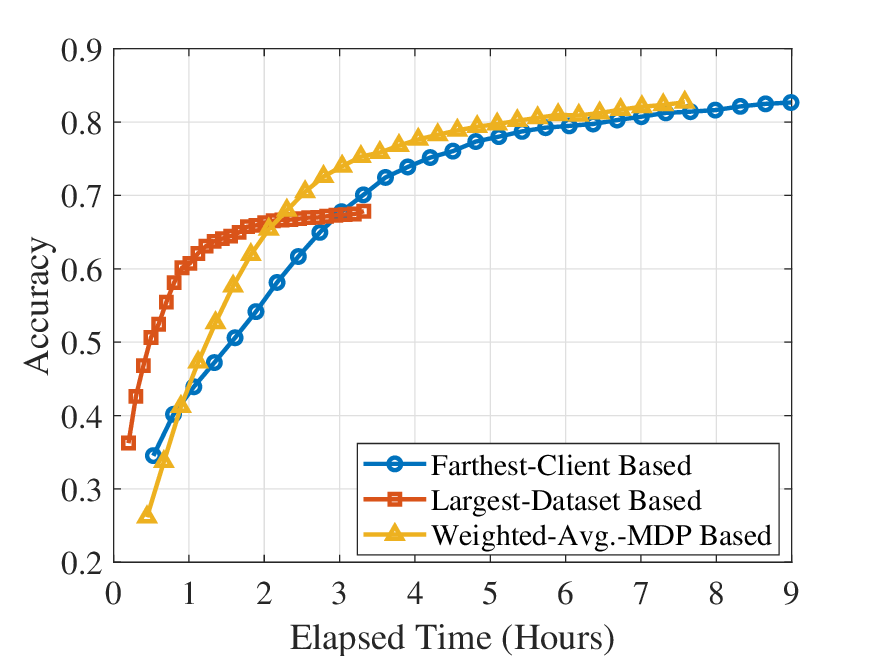}
\caption{Accuracy vs elapsed time.}
\label{fig:2}
\end{figure}

Although Fig.~\ref{fig:1} depicts the accuracies of the schemes as a function of communication rounds, it must be noted that the time to complete one round differs from one LoRa-based scheme to another. This is because each of them uses potentially different SFs for update transfer. To transmit the same update, a higher SF  requires more time than a lower SF. This aspect is illustrated in Fig.~\ref{fig:2}, where we plot the model accuracy after a communication round as a function of the time elapsed since the start of the FL session until the completion of that round. The time needed per round is computed by summing the time needed by the server to transmit the global update and the time needed by the clients to transmit their local update. We observe that the farthest-client strategy incurs the longest completion times. This is because it uses large SFs that are needed to cover the largest server-client distance in the network. By contrast, the largest-dataset strategy achieves the shortest completion times, as it often uses
lower SFs that result in shorter transmissions. However, the time savings come at the cost of significantly lower model accuracy. The weighted-average-MDP strategy performs the best among the three. Its final accuracy is similar to the farthest-client strategy, but the convergence times are lower. For example, to reach 80\% accuracy, it requires about 5.2 hours, whereas the farthest-client method needs about 6.5 hours.

In Table~\ref{tab:tx_times}, we report the average transmission time per communication round for both the client (uplink airtime) and the server (downlink airtime). The primary significance of airtimes is that they are an indirect measure of the energy consumption by the transmitters. We observe that largest-dataset SF allocation provides the lowest airtimes (but at the cost of accuracy penalty as seen earlier), whereas the farthest-client strategy incurs the maximum airtime. The weighted-average-MDP technique requires about 13\% less airtime than the farthest-client strategy. We observe that the average per-client uplink airtimes are slightly lower than downlink airtime. This is because, in those instances where a client does not receive the global update (i.e., encounters more frame erasures than the FEC can correct), the client does not produce a local update for transmission. This reduces the average airtime. In contrast, the server transmits a global update in every communication round.

\begin{table}[!t]
    \caption{Average downlink and per-client uplink airtime (in minutes) per communication round for different SF-allocation strategies}
    \centering
    \begin{tabular}{c|c|c}
        SF-Allocation Strategy & Uplink Airtime & Downlink Airtime \\ \hline
        Farthest-Client Based & 7.58 & 7.82 \\
        Largest-Dataset Based & 2.09 & 2.88 \\
        Weighted-Avg.-MDP Based & 6.57 & 6.87\\
    \end{tabular}
    \label{tab:tx_times}
\end{table}

\section{Conclusion}

We investigated the feasibility of supporting FL for remote IoT over LPWANs, using SHM over LoRa as a representative use case. Our results show that FL over LoRa incurs substantial latency because of long model transmission times. However, if the latency can be tolerated, LoRa-FL can achieve learning accuracies close to those of FL over an ideal (lossless) communication system, provided the transmission parameters are appropriately chosen. We compared three mechanisms for transmission-parameter selection. Our results showed that the weighted-average model-delivery-probability approach balances communication reliability and client contribution, avoiding the excessive conservatism of a worst-case-oriented design while preserving participation from smaller clients that may contain unique classes. Our results provide preliminary design insights for remote IoT deployments where reliable broadband connectivity is unavailable. Future work includes the following:
\begin{itemize}
    \item Extending the framework to larger-scale LPWAN deployments with more clients, diverse network topologies, and different dataset distributions.
    \item Investigating adaptive per-client communication parameter selection, including dynamic spreading-factor, coding-rate, and transmission-power adaptation.
    \item Studying joint optimization of client selection and communication resource allocation to further improve the communication-learning trade-off.
    \item Investigating the proposed communication-aware resource allocation under different data heterogeneity levels and client participation rates.
    \item Evaluating the proposed framework using additional datasets, FL algorithms, and LPWAN technologies, together with experimental validations.
\end{itemize}

\balance
\bibliographystyle{IEEEtran}
\bibliography{refs.bib}

@article{BLN21,
  title={The 1st international project competition for structural health monitoring ({IPC-SHM}, 2020): A summary and benchmark problem},
  author={Bao, Yuequan and Li, Jian and Nagayama, Tomonori and Xu, Yang and Spencer Jr, Billie F and Li, Hui},
  journal={Structural Health Monitoring},
  volume={20},
  number={4},
  pages={2229--2239},
  year={2021},
  publisher={Sage Publications Sage UK: London, England}
}

@inproceedings{SVI16,
  title={Rethinking the inception architecture for computer vision},
  author={Szegedy, Christian and Vanhoucke, Vincent and Ioffe, Sergey and Shlens, Jon and Wojna, Zbigniew},
  booktitle={Proceedings of the IEEE conference on computer vision and pattern recognition},
  pages={2818--2826},
  year={2016}
}

@inproceedings{KKM20,
  title     = {{SCAFFOLD}: Stochastic Controlled Averaging for Federated Learning},
  author    = {Karimireddy, Sai Praneeth and
               Kale, Satyen and
               Mohri, Mehryar and
               Reddi, Sashank and
               Stich, Sebastian and
               Suresh, Ananda Theertha},
  booktitle = {Proceedings of the 37th International Conference on Machine Learning},
  series    = {},
  volume    = {},
  pages     = {5132--5143},
  year      = {2020},
  publisher = {PMLR}
}

@ARTICLE{MSG19,

  author={Mahmood, Aamir and Sisinni, Emiliano and Guntupalli, Lakshmikanth and Rondón, Raúl and Hassan, Syed Ali and Gidlund, Mikael},

  journal={IEEE Transactions on Industrial Informatics}, 

  title={Scalability Analysis of a {LoRa} Network Under Imperfect Orthogonality}, 

  year={2019},

  volume={15},

  number={3},

  pages={1425-1436},

  doi={10.1109/TII.2018.2864681}}

@article{BGG21,
  title={Foundations of population-based {SHM}, {Part I}: Homogeneous populations and forms},
  author={Bull, Lawrence A and Gardner, Paul A and Gosliga, Julian and Rogers, Timothy J and Dervilis, Nikolaos and Cross, Elizabeth J and Papatheou, Evangelos and Maguire, AE and Campos, Carles and Worden, Keith},
  journal={Mechanical systems and signal processing},
  volume={148},
  pages={107141},
  year={2021},
  publisher={Elsevier}
}

@article{PTH25,
  title={{CNN} federated learning for vibration-based damage identification of submerged structure-foundation system},
  author={Pham, Ngoc-Lan and Ta, Quoc-Bao and Huynh, Thanh-Canh and Kim, Jeong-Tae},
  journal={Journal of Civil Structural Health Monitoring},
  volume={15},
  number={7},
  pages={2595--2620},
  year={2025},
  publisher={Springer}
}

@InProceedings{PZZ24,
author="Peng, Zexuan
and Zhang, Zhaohui
and Zhao, Xiaoyan
and Zhang, Tianyao
and Wu, Qi",
editor="Xin, Bin
and Kubota, Naoyuki
and Chen, Kewei
and Dong, Fangyan",
title="Structural Health Monitoring of Similar Gantry Crane Based on Federated Learning Algorithm",
booktitle="Advanced Computational Intelligence and Intelligent Informatics",
year="2024",
publisher="Springer Nature Singapore",
address="Singapore",
pages="75--88"
}

@InProceedings{ASN21,
author="Anaissi, Ali
and Suleiman, Basem
and Naji, Mohamad",
editor="Abreu, Pedro Henriques
and Rodrigues, Pedro Pereira
and Fern{\'a}ndez, Alberto
and Gama, Jo{\~a}o",
title="Intelligent Structural Damage Detection: A Federated Learning Approach",
booktitle="Advances in Intelligent Data Analysis XIX",
year="2021",
publisher="Springer International Publishing",
address="Cham",
pages="155--170",
isbn="978-3-030-74251-5"
}

@ARTICLE{CZC25,
  author={Cheema, Muhammad Asaad and Zohaib Sarwar, Muhammad and Cantero, Daniel and Salvo Rossi, Pierluigi},
  journal={IEEE Internet of Things Journal}, 
  title={Clustered Federated Learning for Population-Based Structural Health Monitoring}, 
  year={2025},
  volume={12},
  number={13},
  pages={24958-24967},
  doi={10.1109/JIOT.2025.3556670}}

@article{YLQ25,
  title={A privacy-preserving framework using federated learning for structural health monitoring with miter gates application},
  author={Yang, Yichao and Lin, Haiwei and Qian, Guofeng and Hu, Zhen and Todd, Michael D},
  journal={Structural Health Monitoring 2025},
  year={2025}
}

@article{AA23,
  author = {Anaissi, Ali and Suleiman, Basem and Alyassine, Widad},
  title = {Personalised federated learning framework for damage detection in structural health monitoring},
  journal = {Journal of Civil Structural Health Monitoring},
  year = {2023},
  volume = {13},
  pages = {295--308},
  doi = {10.1007/s13349-022-00615-y}
}

@ARTICLE{GeR17,  
author={Georgiou, Orestis and Raza, Usman},  
journal={IEEE Wireless Communications Letters},   
title={Low Power Wide Area Network Analysis: Can {LoRa} Scale?},   year={2017},  
volume={6},  
number={2},  
pages={162-165},  
doi={10.1109/LWC.2016.2647247}}

\end{document}